\documentclass{appolb}
\usepackage{epsfig}

\begin{document}
\title{Heavy Flavour Production in \\Two-Photon Collisions at LEP 
\thanks{Presented at the DIS2002 conference in Cracow, Poland; 
to be published in Acta Physica Polonica B}
}
\author{Armin B\"ohrer 
\address{FB Physik, Siegen University,
Emmy-Noether-Campus, 57068~Siegen, Germany}
}
\maketitle
\begin{abstract}
New results from the experiments ALEPH, DELPHI, L3, and OPAL on heavy 
quark production in $\gamma\gamma$ collisions are presented. Inclusive 
charm and bottom production are investigated at LEP~2 energies.  
The total and differential cross sections for charm quarks are now 
measured by all four LEP collaborations, the total bottom by two. 
Charmonia are studied inclusively via the muonic decay of the J$/\psi$ 
and separated for the resolved and diffractive processes. New results 
are available for exclusive production of the $\eta_{\mathrm c}$ meson. 
First searches for exclusive $\eta_{\mathrm b}$ production are presented.
\end{abstract}

\section{Introduction}
\begin{picture}(10,0)(0,0)
\put(345.,250.){SI-2002-3}
\put(345.,240.){June 2002}
\end{picture}
Inclusive heavy flavour production in two-photon collisions is dominated 
by two processes, the direct and single-resolved process. It therefore 
reveals the structure of the photon and is sensitive to its gluon content. 
At LEP~2 energies direct and the single-resolved processes contribute in 
equal shares to heavy flavour final states. The charm cross section is 
about two orders of magnitude larger than the bottom production due to 
the smaller quark mass and higher electric charge. The large quark mass allows 
the production of heavy flavour to be calculated in perturbative QCD, 
where the resolved part also depends on the assumed gluon density of the 
photon.

The exclusive charmonium production has a diffractive contribution at 
low $p_{\mathrm T}^2$ of the vector meson (pomeron exchange) and a resolved 
contribution at high $p_{\mathrm T}^2$ (gluon exchange). 
The resolved production 
of J$/\psi$, when calculated in the nonrelativistic QCD, predicts that 
the colour-octet contribution dominates. Exclusive charmonia and 
bottomonia production at LEP~2 provides a precise tool to test QCD 
at low energies. Their two-photon widths and masses are constraint by 
approaches used in lattice QCD, nonrelativistic QCD and potential models.

In this article we summarize the progress made with respect to last 
year~\cite{dis2001c,dis2001b} in heavy flavour production in two-photon 
collisions at LEP~2: improvements, updates and new results are reviewed.
A recent more general overview on two-photon physics and the interaction 
of the photon may be found in Ref.~\cite{sumascona}.

\section{Inclusive Charm Production}

\subsection{Inclusive D$^*$ Production}
All four LEP experiments~\cite{DstarA,DstarD,DstarL,DstarO} have measured 
the inclusive charm production using all~\cite{DstarA,DstarD,DstarL} or 
two thirds~\cite{DstarO} of their LEP~2 statistics (corresponding to an 
integrated luminosity of $\approx 700\,{\mathrm{pb^{-1}}}$) at 
energies around $\sqrt{s} \approx 200\,{\mathrm{GeV}}$ 
with fully reconstructed D$^*$ mesons in no-tag events. The D$^{*\pm}$ 
mesons are reconstructed via the decay to D$^0\pi^+$, the mass difference 
of D$^*$ and D$^0$ providing a clear signal with small background; up to 
four decay modes of the D$^0$ are considered. 

Three experiments~\cite{DstarA,DstarL,DstarO} provide differential 
distributions in pseudorapidity. The distributions are found to be 
flat in this variable, what is in agreement with the expectation 
for NLO calculations~\cite{nlo} in shape, while in normalisation 
ALEPH and L3 are in agreement with the massive calculation, the OPAL 
data prefer the massless calculation. The distribution in transverse 
momentum to the beam axis as predicted in NLO calculations in the 
massive approach agrees with the data (Figure~\ref{figdstarpt}). 
The data are more consistent 
now as compared to last year's preliminary results~\cite{dis2001c}. 
A small scatter of the data points in the range $2\,{\mathrm{GeV}} < 
p_{\mathrm T} < 3\,{\mathrm{GeV}}$ is still observed, but is within 
the given errors. 

Direct and single-resolved contribution can be separated using the fact that
in the resolved one the remnant jet carries away a part of the invariant
mass available in the $\gamma\gamma$ collision. Two variables have been
used: 1) $x_{\mathrm T} = 2p_{\mathrm T}^{\mathrm{D*}} / W_{\mathrm{vis}}$,
which is the ratio of $p_{\mathrm T}^{\mathrm{D*}}$, a measure for the
invariant mass of the
${\mathrm{c}}\bar{{\mathrm{c}}}$ system, and the visible invariant mass
$W_{\mathrm{vis}}$, a measure for the invariant mass of the $\gamma\gamma$
system; 2) $x^{\mathrm{min}}_{\gamma}$, the
minimum of $x^{\pm}_{\gamma} = \sum_{\mathrm{jets}} (E \pm p_z) /
\sum_{\mathrm{part}} (E \pm p_z)$, a measure
for the fraction of particles, which do not escape in the remnant jet.
The relative contribution $r_{\mathrm{dir}} : r_{\mathrm{res}}$ fitted 
by the three experiments~\cite{DstarA,DstarD,DstarO} are in agreement 
with the prediction of Frixione et al.~\cite{nlo} ($70 : 30$) and 
among themselves regarding the slightly different acceptance ranges 
of the experiments.

\subsection{Cross Section as Function of $W_{\gamma\gamma}$}
The L3 collaboration has measured the charm cross section as
function of the two-photon centre-of-mass energy~\cite{Xsection} 
using the whole LEP~2 data sample. The 
charm-flavoured quarks are identified by their semi-leptonic decays 
to electrons. A parameterization of the
form $\sigma_{\mathrm{tot}} = A s^{\varepsilon} + Bs^{-\eta}$ (Pomeron $+$
Reggeon) describes the data well. The PYTHIA Monte Carlo clearly fails,
predicting only $66\%$ of the total cross section. This may be
partially attributed to next-to-leading order corrections, which are not
included in PYTHIA. The Pomeron slope fitted from the data is steeper than
the rise observed in $\sigma(\gamma\gamma \rightarrow
{\mathrm{q}}\bar{{\mathrm{q}}}{\mathrm{X}})$. NLO order calculations, but using 
a rather small charm quark mass of $1.2\,{\mathrm{GeV}}$, are in very good 
agreement with the data. This also indicates that the resolved contribution, 
the gluon content of the photon, which dominates at high $W_{\gamma\gamma}$, 
is needed to explain the data.

\subsection{Total Charm Cross Section}
\begin{figure}
\begin{minipage}{.50\textwidth}
\epsfxsize=1.00\textwidth
\epsfysize=1.00\textwidth
\epsfbox{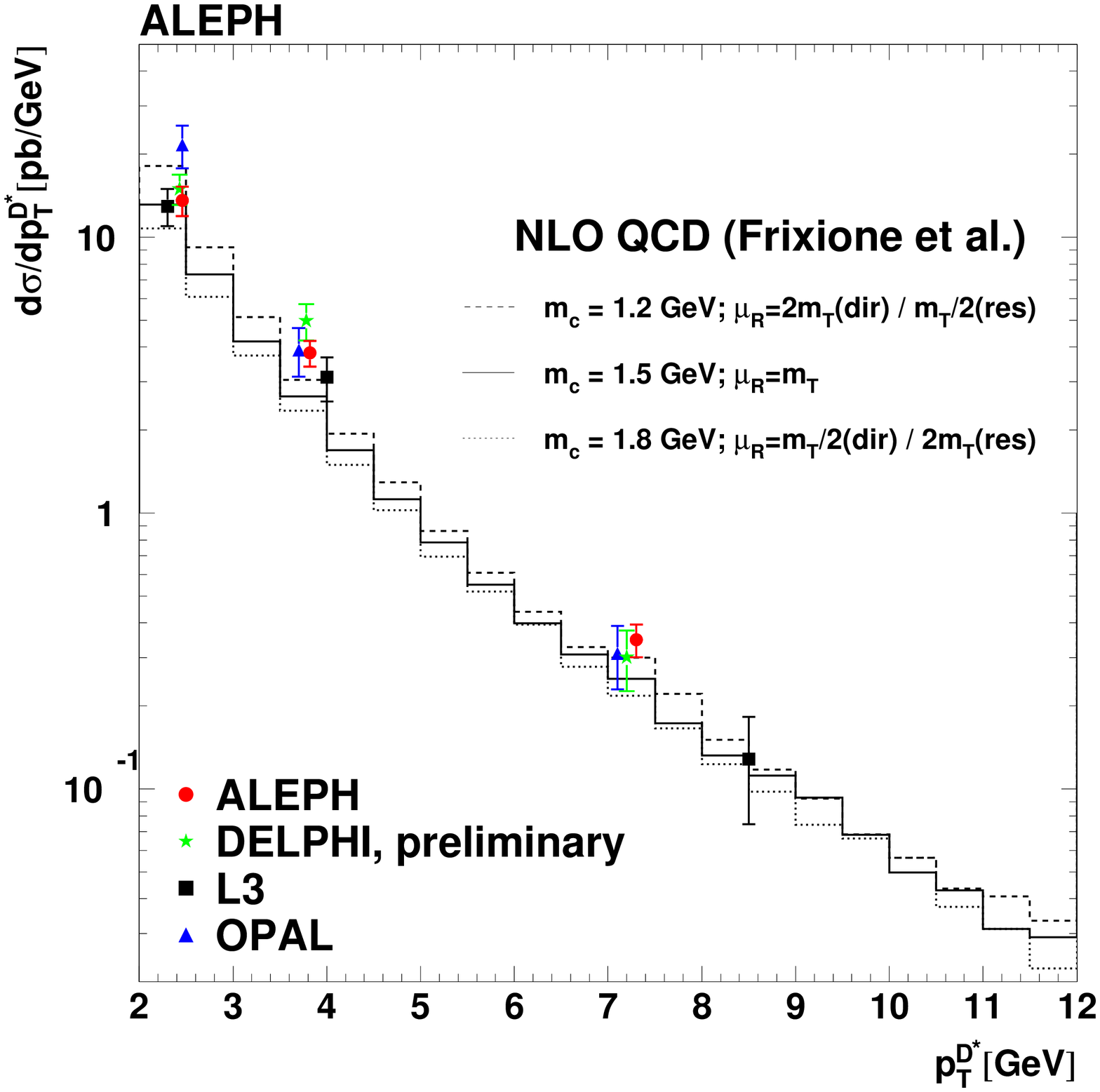}
\caption{Differential D$^*$ cross section in transverse momentum.
\label{figdstarpt}}
\end{minipage}
\begin{minipage}{.50\textwidth}
\epsfxsize=1.00\textwidth
\epsfysize=1.01\textwidth
\epsfbox{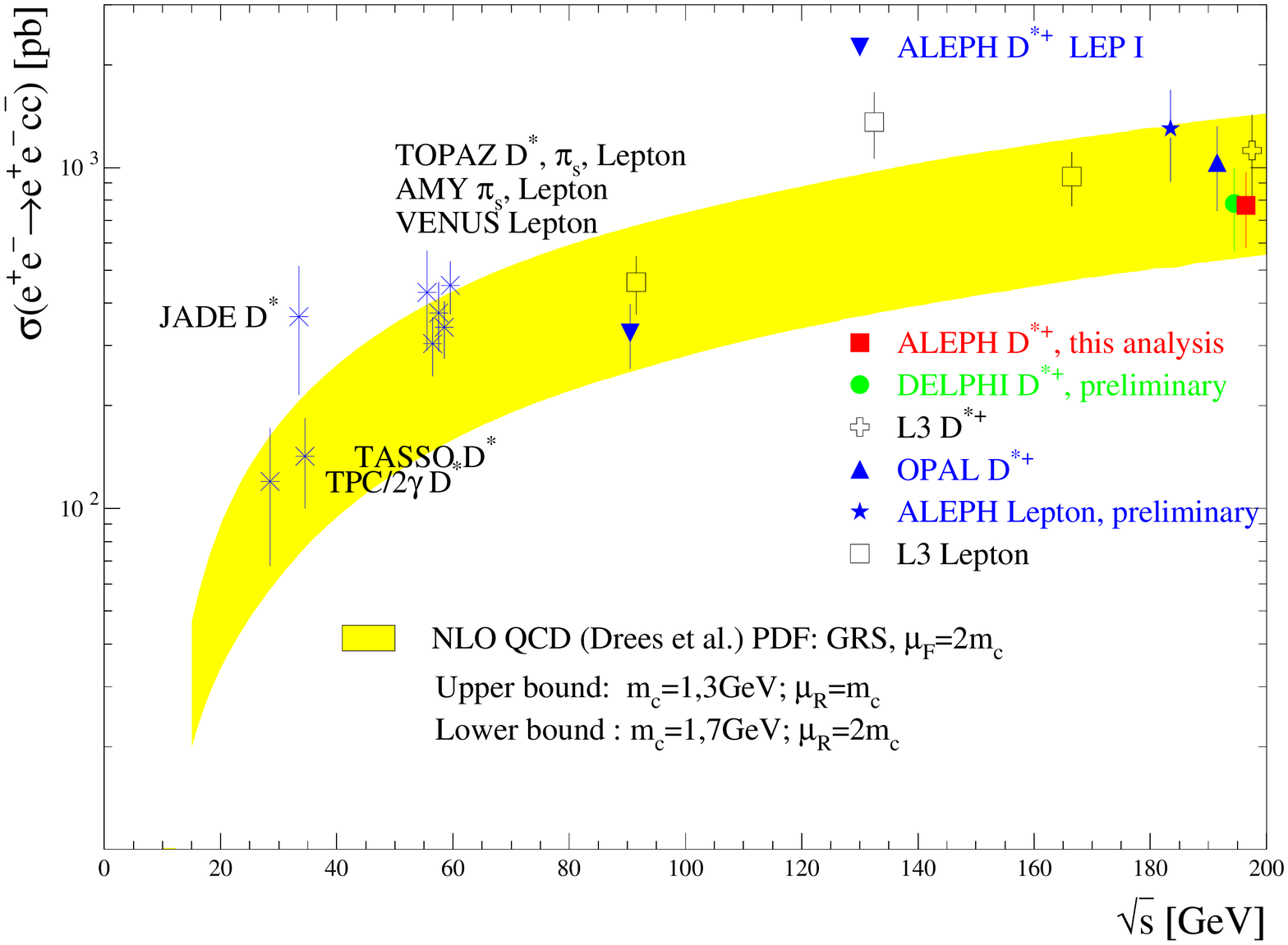}
\caption{Inclusive charm cross section as function of centre-of-mass 
energy.
\label{figcc}}
\end{minipage}
\end{figure}

Extrapolated to the full phase space, the measurements using the D$^*$ as 
the charm tag, can be compared to QCD~\cite{nlo} and other 
measurements~\cite{DstarA}, see Figure~\ref{figcc}. All data are consistent. 
If only the direct contribution would be considered the 
prediction at LEP~2 would be lower by a factor two. It should be 
noted that for the measurements with leptonic final state a light charm 
quark mass is slightly preferred.

\subsection{Charm Structure Function $F_{{\mathrm c},2}^{\gamma}$}
When one of the scattered beam particles is detected, the event can be used
to determine the charm structure function $F^2_{\gamma,{\mathrm {c}}}$. With
60 such single-tagged events with a ${\mathrm{D}}^{*+}$ meson from 
the full LEP~2 statistics, the OPAL
collaboration performed a first measurement in two bins of $x$ with $\langle 
Q^2 \rangle \approx 20\,{\mathrm{GeV}}$~\cite{F2gc}. (See Ref.~\cite{F2gatLEP} 
for a general overview on photon structure functions.) 
The comparison with the calculations
shows that a point-like contribution is not sufficient to describe the data.
A hadron-like part is needed. The data even exceed the models, though the
measurement errors are still too large to be conclusive.

While writing this article, 
OPAL submitted their result~\cite{F2gcnew} for publication now with a fitted 
number of 55 ${\mathrm{D}}^{*+}$ mesons, reducing somewhat the discrepancy 
between data and the calculations at low $x$.

\section{Inclusive Bottom Production}
Open bottom production is measured by the L3 and the OPAL
col\-la\-bo\-ra\-tions~\cite{bottom} at LEP~2 energies using 
an integrated luminosity of
$400\,{\mathrm{pb^{-1}}}$. Their analysis procedures exploit the
fact, that the momentum as well as the transverse momentum of leptons
with respect to the closest jet is higher for muons and electrons from bottom
than from background, which is mainly charm. Therefore, leptons with momenta
of more than $2\,{\mathrm{GeV}}$ are selected and their momentum distribution
with respect to the closest jet (obtained with the JADE jet-algorithm in L3
and KTCLUS in OPAL, while in both experiments the lepton was excluded,
when defining the jet) is investigated.

Similar to the studies in charm production, the bottom quarks produced
in direct and single resolved events show a different behaviour in the
transverse momentum distribution. The variable $x_{\mathrm T}^{\mu} =
2p_{\mathrm T}^{\mu}/W_{\mathrm{vis}}$ is well suited to demonstrate the
need for both contributions: the single resolved part at low
$x_{\mathrm T}^{\mu}$ and the direct part at high $x_{\mathrm T}^{\mu}$. 
The agreement between data and Monte Carlo
simulation is very good.

\begin{figure}
\begin{minipage}{.50\textwidth}
\epsfxsize=1.00\textwidth
\epsfysize=1.00\textwidth
\epsfbox{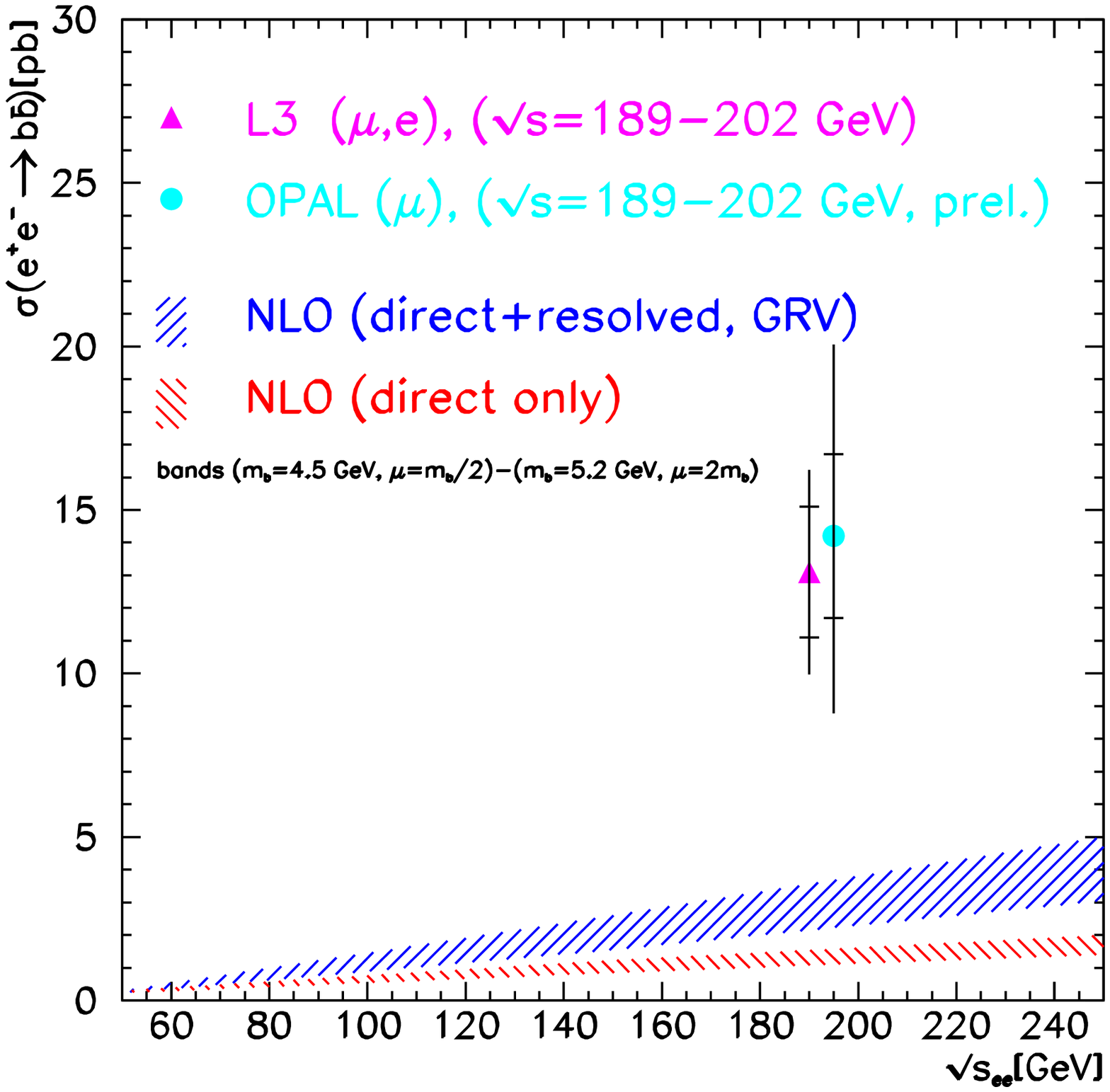}
\caption{Inclusive bottom cross section.
\label{figbb}}
\end{minipage}
\begin{minipage}{.50\textwidth}
\includegraphics[width=1.00\textwidth,clip]{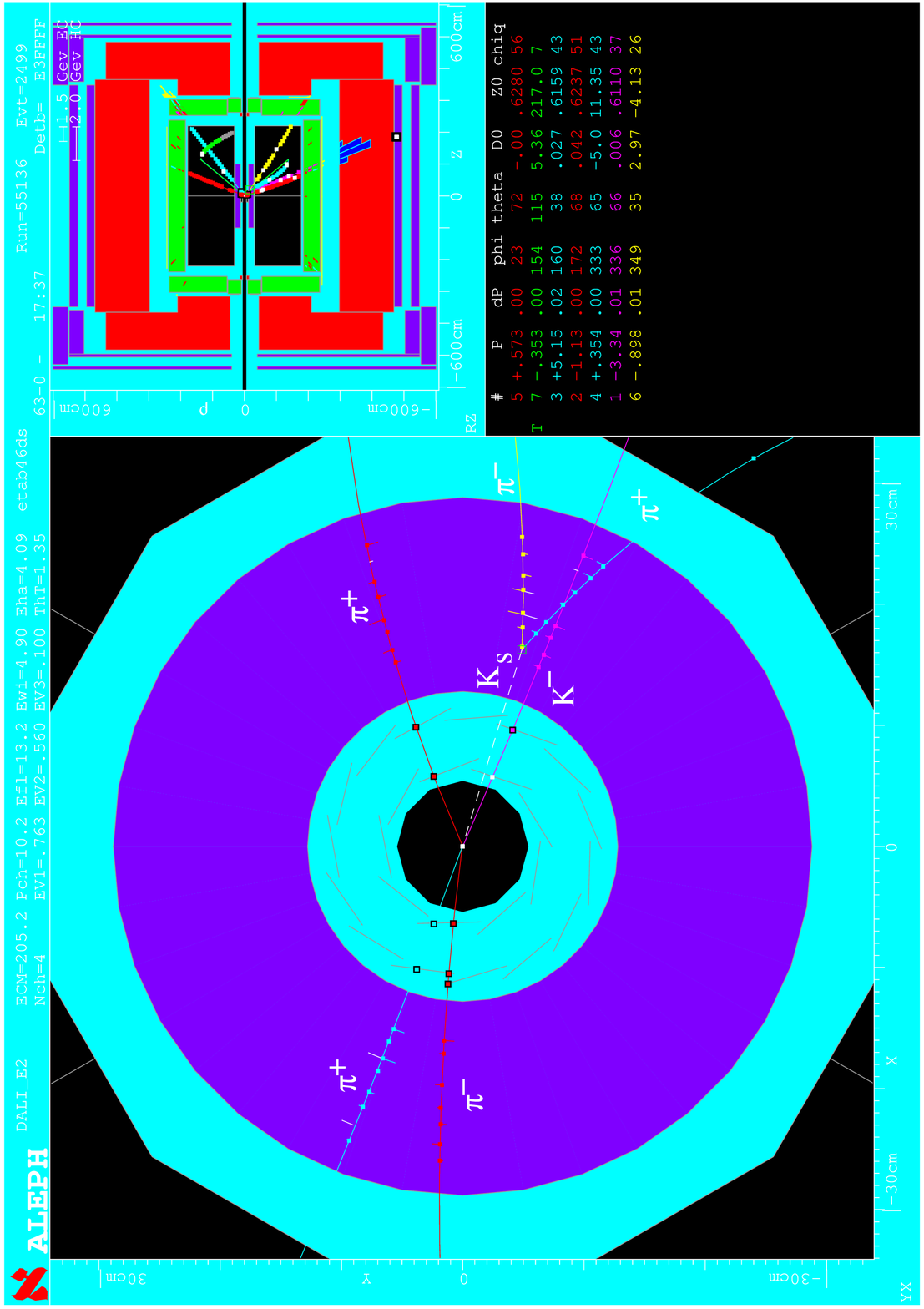}
\caption{candidate event.
\label{figetab}}
\end{minipage}
\end{figure}

The total cross section measurements for open bottom production
are summarized in Figure~\ref{figbb}. The results are compared to
NLO calculations~\cite{nlo}. The calculations
underestimate the data by a factor three corresponding to three to four 
standard deviations. See also remarks and discussions in the presentation 
in Ref.~\cite{bottomtalk}.

\section{Production of Charmonia and Bottomonia}

\subsection{Inclusive J$/\psi$}
A first study was performed by DELPHI of inclusive J$/\psi$ production in 
$\gamma\gamma$ collisions at LEP~2 energies~\cite{Jpsi}. A 
clean signal of $36\pm 7$ 
events for $\gamma\gamma \rightarrow$J$/\psi +$X is seen, where the 
J$/\psi$ decays to a muon pair and where X denotes 
at least two tracks seen in the detector. From a fit of the direct 
and resolved contribution, as taken from the PYTHIA simulation, a value 
of $74 \pm 22\%$ is extracted to originate from the resolved process: a
clear indication for the gluon in the photon. In a recent paper the octet 
production of J$/\psi$ in association with jets has been 
discussed~\cite{JpsiKla}. It was argued that the singlet production is 
not sufficient in order to describe the data, but that the octet 
production is needed for an agreement with the DELPHI measurements. 
See also Ref.~\cite{JpsiLee} for a discussion of octet production.

\subsection{Exclusive $\eta_{\mathrm c}$}
The formation of the $\eta_{\mathrm c}$ in exclusive production in 
two-photon production is a good test of QCD (See Ref.~\cite{braccini} for 
a short summary of the present status of studies of exclusive particle 
production in two-photon events). A preliminary study of 
$\eta_{\mathrm c}$ at LEP~2 has been contributed to this 
conference~\cite{etac}. Nice signals are found in the decay modes 
of the $\eta_{\mathrm c}$ to $\pi^+\pi^-$K$^+$K$^-$, K$^+$K$^-$K$^+$K$^-$, 
and K$_{\mathrm {S}}$K$^+\pi^-$. No signal, however, is seen in the 
$\pi^+\pi^-\pi^+\pi^-$, while for the latter it is also expected, when 
the branching fraction of Ref.~\cite{RPP} are implied. An upper limit for 
this channel is given. For the other three a two-photon width is obtained 
somewhat higher than the world averages~\cite{RPP} and recent measurements 
at LEP and other colliders~\cite{braccini}.

\subsection{Exclusive $\eta_{\mathrm b}$}
As reported at last year's DIS conference in Bologna~\cite{dis2001b}, 
the ALEPH experiment has started a search for the still undiscovered
$\eta_{\mathrm b}$ pseudoscalar meson (See Figure~\ref{figetab} for 
a candidate event). The search has recently been published~\cite{etabA}; 
a preliminary result of a search has now also been reported by 
L3~\cite{etabL}. 

Various predictions exist for
the mass of the $\eta_{\mathrm b}$, e.g., from potential models, pQCD,
NRQCD, and lattice calculations. While the production can reliably be
estimated the branching ratios of the meson have to be guessed. 
An estimate based on MLLA combined with LPHD and using isospin invariance 
has been proposed~\cite{etabascona} and used by ALEPH. 
The efficiencies for the decay modes of the $\eta_{\mathrm b}$ under study 
by ALEPH (four charged particles; or six charged particles) and L3 
(two charged particle plus neutral pion or eta; or four charged particles; or 
six charged particles) are around 13\% and 4\%, respectively. The experiments 
would expect about one to two events. The background is estimated 
to be about one event. ALEPH observes one, L3 observes three events in 
the signal region from $9.0\,{\mathrm{GeV}}$ to $9.8\,{\mathrm{GeV}}$. 
Candidate masses are $9.30 \pm 0.03\,{\mathrm{GeV}}$ and 
$9.49 \pm 0.30\,{\mathrm{GeV}}$, $9.14 \pm 0.30\,{\mathrm{GeV}}$, 
$9.70 \pm 0.30\,{\mathrm{GeV}}$, respectively. Limits on 
$\Gamma_{\gamma\gamma}(\eta_{\mathrm b})$ and 
$\Gamma_{\gamma\gamma}(\eta_{\mathrm b})\times$BR have been given, e.g., 
$\Gamma_{\gamma\gamma}(\eta_{\mathrm b})\times$BR(4cha)$<48\,{\mathrm{eV}}$ 
and 
$\Gamma_{\gamma\gamma}(\eta_{\mathrm b})\times$BR(6cha)$<132\,{\mathrm{eV}}$ 
by ALEPH.

\section{Summary}
We summarized the investigations of heavy flavour production 
in two-photon collisions 
at LEP. The inclusive charm production has been studied and QCD predictions 
and experiment are found in agreement for differential distributions such as 
in pseudorapidity and transverse momentum and total cross section. The 
dependence on $W_{\gamma\gamma}$ is reproduced by QCD, which requires gluon in 
the photon. This is also proven by the explicit measurement of the fraction 
of direct an resolved contribution in no-tag events. The charm structure 
function $F_{\mathrm {c,2}}^{\gamma}$ shows some deviations, but more 
statistics is needed. 

The inclusive bottom cross section shows a serious problem. The total 
cross section as predicted by NLO QCD-calculations is too low by a factor 
three.

Newer measurements on inclusive charmonia clearly show that the colour octet 
production is needed. While most measurements of exclusive charmonia are 
in agreement with the world averages some inconstistencies arose, such as in 
the decay of the $\eta_{\mathrm c}$ to four charged pions. 
The searches of the $\eta_{\mathrm b}$ in exclusive production from two 
photons look promising, but other experiments are needed for discovery.


\newpage

\begin{thebibliography}{99}

\bibitem{dis2001c}
S.~Vlachos, {\it Charm Production in Two-Photon Collisions at LEP}, 
to appear in proceedings of DIS 2001, Bologna, Italy (2001), 
ed.\ by G.~Bruni, G.~Iacobucci, R.~Nania, World Sci., Singapore, 2001.
\bibitem{dis2001b}
A.~B\"ohrer, {\it Bottom Production in Two-Photon Collisions at LEP}, 
to appear in proceedings of DIS 2001, Bologna, Italy (2001), 
ed.\ by G.~Bruni, G.~Iacobucci, R.~Nania, World Sci., Singapore, 2001.
\bibitem{sumascona}
A.~B\"ohrer and M.~Krawczyk, {\it Conference Summary}, 
to appear in proceedings of PHOTON 2001, Ascona, Switzerland (2001),
ed.\ by M.~Kienzle, World Sci., Singapore, 2001.
\bibitem{DstarA}
ALEPH Collaboration, paper in preparation;\\
A.B.~Ngac, private communications.
\bibitem{DstarD}
DELPHI Collaboration, conference contribution (PHOTON 2001).
\bibitem{DstarL}
L3 Collaboration, {\it CERN-EP/2002-012}.
\bibitem{DstarO}
OPAL Collaboration, {\it OPAL-Note PN 453} (2000).
\bibitem{nlo}
M.~Drees, M.~Kr\"amer, J.~Zunft and P.M.~Zerwas,
{\it Phys. Lett.} B {\bf 306}, 371 (1993);\\
S.~Frixione, M.~Kr\"amer and E.~Laenen,
{\it hep-ph/9908483}, {\it Nucl. Phys.} B {\bf 571}, 169 (2000).
\bibitem{Xsection}
L3 Collaboration, {\it Phys. Lett.} B {\bf 514}, 19 (2001).
\bibitem{F2gc}
OPAL Collaboration, {\it OPAL-Note PN 490} (2001).
\bibitem{F2gcnew}
OPAL Collaboration, {\it CERN-EP/2002-31}.
\bibitem{F2gatLEP}
M.~Przybycien, these proceedings.
\bibitem{bottom}
L3 Collaboration, {\it Phys. Lett.} B {\bf 503}, 10 (2001);\\
OPAL Collaboration, {\it OPAL-Note PN 455} (2000).
\bibitem{bottomtalk}
A.~Szczurek, these proceedings;\\
L.~Lonnblad, J.~Repond, M.~Zielinski, these proceedings.
\bibitem{Jpsi}
DELPHI Collaboration, {\it DELPHI-Note 2002-013} (2002).
\bibitem{JpsiKla}
M.~Klasen {\it et al.}, {\it DESY 01-202, hep-ph/0112259} (2001).
\bibitem{JpsiLee}
J.~Lee, these proceedings.
\bibitem{braccini}
S.~Braccini, {\it Resonances and Exclusive Channels: An Experimenter's 
Summary}, 
to appear in proceedings of PHOTON 2001, Ascona, Switzerland (2001),
ed.\ by M.~Kienzle, World Sci., Singapore, 2001.
\bibitem{etac}
DELPHI Collaboration, conference contribution (PHOTON 2001);\\
A.~Oblakowska-Mucha , private communications.
\bibitem{RPP}
Particle Data Group,
{\em Review of Particle Physics}, {\it Eur. Phys. J.} C {\bf 15}, 1 (2000).
\bibitem{etabA}
ALEPH Collaboration, {\it Phys. Lett.} B {\bf 530}, 56 (2002).
\bibitem{etabL}
L3 Collaboration, {\it L3-Note 2736} (2002).
\bibitem{etabascona}
A.~B\"ohrer, {\it Search for the $\eta_b$ Meson}, 
to appear in proceedings of PHOTON 2001, Ascona, Switzerland (2001),
ed.\ by M.~Kienzle, World Sci., Singapore, 2001.
\end{thebibliography}
\end{document}